\documentclass[a4paper,11pt]{article}
\usepackage{pos}
\usepackage{slashed}
\usepackage{braket}
\usepackage[most]{tcolorbox}
\usepackage[normalem]{ulem}

\title{
Soft-overlap contribution to $B_c \to \eta_c$ form factors: diagrammatic resummation of double logarithms
}
\ShortTitle{$B_c \to \eta_c$ form factors in the double logarithmic approximation}

\author[a]{Guido Bell}
\author*[b]{Philipp Böer}
\author[a]{Thorsten Feldmann}
\author[a]{Dennis Horstmann}
\author[a]{Vladyslav Shtabovenko}

\affiliation[a]{Theoretische Physik 1, Center for Particle Physics Siegen,\\
Universit\"at Siegen, 57068 Siegen, Germany}
\affiliation[b]{PRISMA\textsuperscript{+} Cluster of Excellence \& Mainz Institute for Theoretical Physics,\\
Johannes Gutenberg Universit\"at, 55099 Mainz, Germany}

\emailAdd{bell@physik.uni-siegen.de}
\emailAdd{pboeer@uni-mainz.de}
\emailAdd{thorsten.feldmann@uni-siegen.de}
\emailAdd{horstmann@physik.uni-siegen.de}
\emailAdd{shtabovenko@physik.uni-siegen.de}

\abstract{
Using diagrammatic resummation techniques, we investigate the double-logarithmic series of the ``soft-overlap'' contribution to $B_c \to \eta_c$ transition form factors at large hadronic recoil, assuming the scale hierarchy $m_b \gg m_c \gg \Lambda_{\rm QCD}$. In this case, the hadronic bound states can be treated in the non-relativistic approximation and the relevant hadronic matrix elements can be computed perturbatively. This setup defines one of the simplest examples to study the problem of endpoint singularities appearing in the factorization of exclusive $B$-decay amplitudes. We find that the leading double logarithms arise from a peculiar interplay of soft-quark ``endpoint logarithms'' from ladder diagrams with energy-ordered spectator-quark propagators, as well as standard Sudakov-type soft-gluon corrections. We elucidate the all-order systematics, and show that their resummation proceeds via a novel type of integral equations.
The current status of the calculation, which includes all double logarithms in the Abelian limit, is reported.
}

\FullConference{16th International Symposium on Radiative Corrections: Applications of Quantum Field Theory to Phenomenology 
(RADCOR2023)\\
28th May - 2nd June, 2023\\
Crieff, Scotland, UK\\[2ex]

SI-HEP-2023-19, P3H-23-058, MITP-23-043}

\begin{document}
\maketitle

\section{Introduction}
\label{sec:introduction}

\noindent 
Decays of $B$ mesons provide an excellent laboratory for measuring fundamental parameters of the Standard Model, or to look for yet unknown particles and interactions. Their theoretical description relies to a vast extent on the relatively large mass of the $b$-quark as compared to the strong interaction scale $\Lambda_{\rm QCD}$, which allows one to expand the decay amplitudes in the small ratio \mbox{$\Lambda_{\rm QCD}/m_b \ll 1$}. This expansion is well-understood for processes that admit a local operator product expansion, but the situation is strikingly different for decays into light and energetic hadrons. In particular, a framework to systematically include power corrections to the QCD factorization approach~\cite{Beneke:1999br,Beneke:2000ry,Beneke:2001ev}
is currently unknown.

In exclusive heavy-to-light transitions, the relevant non-perturbative dynamics is captured by matrix elements of light-ray operators, so-called light-cone distribution amplitudes (LCDAs), that are convoluted with a perturbative hard-scattering kernel. While this picture was first developed using diagrammatic QCD factorization techniques, power corrections in the heavy-quark expansion can be addressed more systematically with methods from Soft-Collinear Effective Theory~\cite{Bauer:2000yr,Bauer:2001yt,Beneke:2002ph,Beneke:2002ni}. Among the many technical challenges that arise at subleading power, the appearance of endpoint-divergent convolution integrals is certainly the biggest hurdle to overcome. Despite recent progress in their consistent treatment, see e.g.~\cite{Boer:2018mgl,Liu:2019oav,Beneke:2020ibj,Bell:2022ott,Feldmann:2022ixt,Cornella:2022ubo,Hurth:2023paz}, based on refactorization ideas~\cite{Boer:2018mgl,Liu:2019oav}, endpoint-divergent convolutions in hard exclusive processes remain elusive due to their nested structure~\cite{Boer:2018mgl,Bell:2022ott}. This was illustrated for a basic QED process in~\cite{Bell:2022ott}, and in the current work we study the all-order structure of the associated double logarithmic corrections to heavy-to-light transition form factors at large hadronic recoil. To this end, we consider $B_c \to \eta_c$ form factors in the non-relativistic (NR) approximation as a perturbative template case, employing the scale hierarchy $m_b \gg m_c \gg \Lambda_{\rm QCD}$.

\section{Setup and notation}
\label{sec:notation}

\noindent 
Bottom and charm quarks are sufficiently heavy to justify the treatment of $B_c$ and  $\eta_c$ mesons as NR bound states. In this limit, the $B_c$ meson consists of a heavy $b$-quark with momentum \mbox{$p_b^\mu = m_b v^\mu$}, and a soft spectator $\bar{c}$-quark with momentum $\ell^\mu = m_c v^\mu$. Its spinor degrees-of-freedom are encoded in the Dirac projector $\frac12 (1+ \slashed{v})\gamma_5$, with $v^\mu$ being the meson's four-velocity ($v^2 = 1$). Correspondingly, the $\eta_c$ is described by two on-shell quarks with collinear momentum  $p^\mu = m_c v'^\mu$, and its Dirac projector is $\frac12 (1-\slashed{v}')\gamma_5$. The meson masses $m_B$ and $m_\eta$, and momenta $p_B^\mu$ and $p_\eta^\mu$, are simply the sum of the corresponding quark masses and momenta, respectively. As the quark masses provide a physical infrared regulator, exclusive $B_c \to \eta_c$ transition matrix elements are computable in perturbation theory in this approximation~\cite{Bell:2005gw,Bell:2006tz,Boer:2018mgl}.  

At large hadronic recoil, the velocities of the mesons are expressed in standard light-cone coordinates as $v^\mu = \frac12 n^\mu + \frac12 \bar{n}^\mu$ and $v'^\mu \simeq \gamma n^\mu + \frac{1}{4\gamma} \bar{n}^\mu$, with $n^2=\bar n^2=0$ and $n\cdot \bar n=2$. Here $\gamma \equiv v\cdot v' = \mathcal{O}(m_b/m_c)$ denotes the parametrically large boost of the $\eta_c$ meson in the $B_c$ rest frame. In the following, we consider a transition form factor defined by the hadronic matrix element
\begin{align}
    F(\gamma) \equiv \frac{1}{2E_\eta} \bra{\eta_c(p_\eta)} \big( \bar{c} \Gamma b\big)(0) \ket{B_c(p_B)} \,, \qquad \text{with} \qquad \Gamma = \frac{\slashed{\bar{n}} \slashed{n}}{4} \,,
\end{align}
and the large energy of the $\eta_c$ meson $E_\eta = \gamma m_\eta = \mathcal{O}(m_B)$. The Dirac structure $\Gamma$ is chosen to project out the so-called soft-overlap contribution, whose factorization in terms of LCDAs is known to be spoilt by endpoint-divergent convolution integrals~\cite{Beneke:2000wa,Beneke:2003pa,Lange:2003pk}.

\section{Fixed-order analysis}
\label{sec:FO}

\begin{figure}[t]
    \centering
   \includegraphics[width=0.38\textwidth]{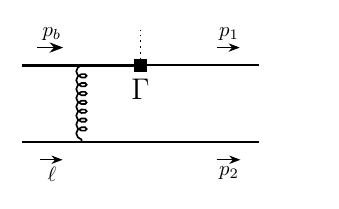} \hspace{1cm}
    \includegraphics[width=0.38\textwidth]{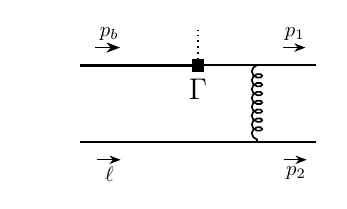}
    \vspace{-0.2cm}
    \caption{Tree-level contributions to the non-relativistic transition form factors.
\label{fig:LOdiagrams}}
\end{figure}

\noindent 
The two diagrams shown in Fig.~\ref{fig:LOdiagrams} yield the tree-level expression for the form factor
\begin{align}
\label{eq:xiLO}
 F^{(0)}(\gamma) = \xi_0 \;\frac{2+\bar{u}_0}{\bar{u}_0^3} \,, \qquad \text{with} \qquad \xi_0 = \frac{\alpha_s C_F}{4\pi} \frac{\pi^2 f_{\eta} f_B m_B}{N_c E_\eta^2 m_\eta} \,,
\end{align}
and where $f_{\eta}$ and $f_B$ are the meson decay constants.\footnote{Distinguishing the QCD decay constants from their NR or static limits is a single logarithmic effect, which is irrelevant for our analysis.} For convenience, we distinguish the spectator-quark mass $m_2$ from the active-quark mass $m_1$, which introduces the mass ratios $u_0 = m_1/m_\eta$ and $\bar{u}_0 = 1-u_0 = m_2/m_\eta$, with $u_0 = \bar{u}_0 = 1/2$ for the $\eta_c$ meson. The momenta of the light quarks are then given by $p_i^\mu=m_i v'^\mu$ for $i=1,2$ and $\ell^\mu=m_2 v^\mu$, see Fig.~\ref{fig:LOdiagrams}. We remark that, when working in light-cone gauge $\bar{n} \cdot A = 0$, only the second diagram contributes to the soft-overlap form factor, as energetic gluons couple to the heavy $b$-quark only via eikonal Wilson-line interactions. This will be an important observation for the analysis of higher-order corrections described below.

Our goal consists in resumming all double logarithmic corrections of the form $\alpha_s^n \ln^{2n}(2\gamma)$ to the form factor $F(\gamma)$. To this end, we expand the form factor in powers of $\frac{\alpha_s}{4\pi}$, and indicate the loop-order with a superscript $(n)$. Before examining the all-order structure, it is instructive to discuss in some detail how the double logarithms arise in fixed-order perturbation theory at next-to-leading order (NLO) and next-to-next-to-leading order (NNLO).

\subsection{Next-to-leading order}
\label{subsec:NLO}

\begin{figure}[t]
    \centering
   \includegraphics[width=0.23\textwidth]{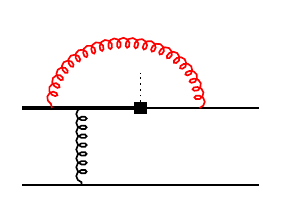}
   \includegraphics[width=0.23\textwidth]{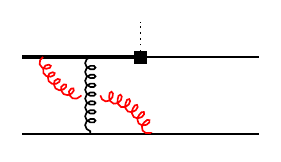}
   \includegraphics[width=0.23\textwidth]{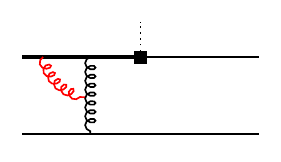}   
   \includegraphics[width=0.23\textwidth]{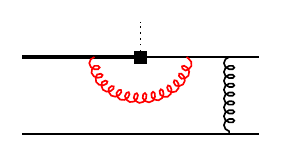} \\ 
   \includegraphics[width=0.23\textwidth]{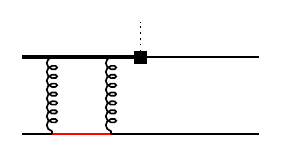}
   \includegraphics[width=0.23\textwidth]{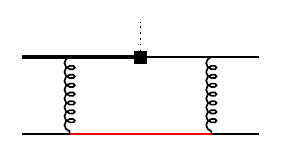}
   \includegraphics[width=0.23\textwidth]{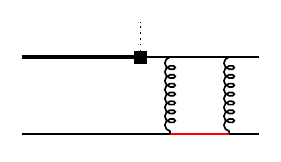}
   \includegraphics[width=0.23\textwidth]{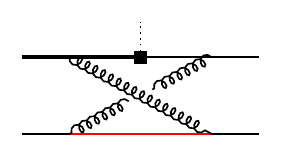}
    \caption{At one-loop order and in Feynman gauge, all double-logarithms arise from the depicted diagrams in configurations with on-shell soft gluon or quark propagators (red), and eikonal off-shell propagators (black).
\label{fig:NLO}}
\end{figure}

\noindent 
From now on we use the symbol $\simeq$ to indicate the double-logarithmic approximation. A straight-forward calculation gives at one-loop order~\cite{Bell:2006tz,Boer:2018mgl}
\begin{align}
\label{eq:NLODL}
     F^{(1)}(\gamma)  \simeq \xi_0 L^2 \left( C_F \frac{1+2\bar{u}_0}{\bar{u}_0^3} - \frac{C_A}{2\bar{u}_0^3}\right) \,, \qquad \text{with} \qquad L \equiv \ln(2\gamma) \,.
\end{align}
This result can be reproduced in an economic way by considering kinematic limits of the subset of Feynman diagrams shown in Fig.~\ref{fig:NLO} (in Feynman gauge). It turns out that the form factor features two sources of double logarithms. First, from standard soft-gluon configurations of the diagrams in the first line of Fig.~\ref{fig:NLO}, and second, from soft spectator-quark propagators in the diagrams shown in the second line. In the following we explain how to extract the relevant double logarithms from one representative of each class.

In the soft-gluon limit $k^\mu \to 0$, the last diagram in the first line of Fig.~\ref{fig:NLO} factorizes into
\begin{equation}
\label{eq:softgluonfactorization}
  \vcenter{\hbox{\vspace{+0.2cm} \includegraphics[width=0.2\textwidth]{figs/softglue3.pdf}}} \simeq \vcenter{\hbox{\vspace{+0.2cm} \includegraphics[width=0.2\textwidth]{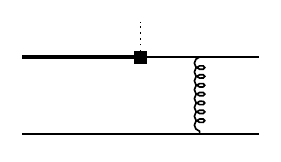}}} \times \, \frac{\alpha_s C_F}{4\pi} \, (-16i \pi^2) \int \!\! \frac{d^4k}{(2\pi)^4} \, \frac{1}{k^2 \, (v \cdot k) \, (\ell_+ + k_+)} \,,
\end{equation}
where we kept the $i\epsilon$-prescription of the propagators implicit, and wrote $q^\mu = q_- \frac{n^\mu}{2}+ q_\perp^\mu + q_+ \frac{\bar{n}^\mu}{2}$ for the light-cone decomposition of an arbitrary four-vector $q^\mu$. The integral in~\eqref{eq:softgluonfactorization} is double-logarithmic if the two quark propagators become eikonal along the light-cone directions,  
\begin{align}
    (m_b v - k)^2 - m_b^2 \approx -2 m_b v\cdot k &\stackrel{!}{\simeq} - m_b \, k_- & &\Rightarrow \quad k_- \gg k_+ \,, \\
    (p_\eta-\ell-k)^2-m_1^2 \approx 
    -p_{\eta-} (\ell_+ +k_+) &\stackrel{!}{\simeq} 
    - p_{\eta-} k_+ & &\Rightarrow \quad 
    p_{\eta-}
    \gg k_- \quad\text{and}\quad k_+ \gg \ell_+ \,, \nonumber 
\end{align}
which restricts the integration domain to a triangle. The soft-gluon propagator effectively goes on-shell, \mbox{$1/k^2 \to -2\pi i \delta(k_+ k_- + k_\perp^2)$}, and after integrating the transverse momenta one finds 
\begin{align}
  \vcenter{\hbox{\vspace{+0.2cm} \includegraphics[width=0.2\textwidth]{figs/softglue3.pdf}}} &\simeq \vcenter{\hbox{\vspace{+0.2cm} \includegraphics[width=0.2\textwidth]{figs/LO2.pdf}}} \times \, \frac{\alpha_s C_F}{4\pi} (-2) \int_{\ell_+}^{p_{\eta-}} \!\! \frac{dk_-}{k_-} \int_{\ell_+}^{k_-} \!\! \frac{dk_+}{k_+} \,. 
\end{align}
At the double-logarithmic level the integral evaluates to $\frac12 L^2$, and unsurprisingly the contribution from this diagram is the tree-level graph multiplied by the universal one-loop Sudakov logarithm from the cusp at the heavy-to-light vertex. After summing all soft-gluon contributions from Fig.~\ref{fig:NLO}, one recovers the full LO contribution from~\eqref{eq:xiLO} multiplied by the  one-loop Sudakov factor.  

We next consider the leftmost box diagram in the second line of Fig.~\ref{fig:NLO}. Here a double logarithm arises from an on-shell soft spectator-quark propagator, $1/(k^2-m_2^2) \to -2 \pi i \delta(k_+ k_- + k_\perp^2 - m_2^2)$, with $k$ now being the soft-quark momentum. Integrating the transverse components then results in the phase-space constraint \mbox{$\theta(k_+ k_- - m_2^2)$}.  Again, all remaining propagators become eikonal, but now the numerator of the spectator-quark propagator provides an additional factor $k_-$ to guarantee logarithmic sensitivity of both longitudinal integrations. Expanding the propagator-denominators in the double-logarithmic approximation now gives the integration boundaries $\ell_-<k_-<p_{2-}$ and $\ell_+>k_+>p_{2+}$, and one obtains for the respective double logarithm of this diagram
\begin{align}
\label{eq:NLOsoftquarkexample}
  \frac{\alpha_s C_F}{4\pi} \, \xi_0 \,\frac{2}{\bar{u}_0^3} \, \int_{\ell_-}^{p_{2-}} \! \frac{dk_-}{k_-} \int_{p_{2+}}^{\ell_+} \! \frac{dk_+}{k_+} \, \theta(k_+ k_- - m_2^2) = \frac{\alpha_s C_F}{4\pi} \, \xi_0 L^2 \frac{1}{\bar{u}_0^3} \,.  
\end{align}
Proceeding similarly for all displayed diagrams yields
\begin{align}
\label{eq:NLODLgluequark}
F^{(1)}(\gamma) \simeq 
\underbrace{- C_F L^2 \times F^{(0)}(\gamma) }_{\text{soft-gluon corrections}} \quad + \quad  \underbrace{\xi_0 L^2 \left( 3 C_F \frac{1+\bar{u}_0}{\bar{u}_0^3} - \frac{C_A}{2\bar{u}_0^3}\right)}_{\text{soft-quark corrections}} \,, 
\end{align}
which adds up to the result in~\eqref{eq:NLODL}. We remark that only the soft-gluon logarithms are proportional to the tree-level form factor as they arise from eikonal couplings, in contrast to the soft-quark configurations. 

The observed origins of large logarithms are related to the factorization of the soft-overlap form factor in SCET. Whereas soft-gluon logarithms arise from standard cusp-induced contributions, soft-quark logarithms result from an overlapping spectator-quark propagator, and are associated with \emph{endpoint-divergent} convolutions of the $B_c$ and $\eta_c$ LCDAs. We hence refer to them as ``endpoint-logarithms''. Presently, neither their higher-order structure, nor a resummation program based on renormalization-group equations in an effective-field-theory framework has been known. 

\subsection{Next-to-next-to-leading order}
\label{subsec:NNLO}

\noindent 
We now elaborate on the leading logarithms at two-loop order in the Abelian limit ($C_A=0$). Soft-gluon corrections are known to exponentiate, and hence they contribute as \mbox{$+\frac12 C_F^2 L^4 \times F^{(0)}(\gamma)$} to the form factor. More interestingly, they now also modify the soft-quark ``endpoint'' configurations in two ways. The first piece is the product of the universal one-loop Sudakov factor and the one-loop soft-quark contribution from~\eqref{eq:NLODLgluequark}. However, also the \emph{integrands} of the expressions of the type discussed in~\eqref{eq:NLOsoftquarkexample} receive a universal double logarithm that depends on the longitudinal integration variables $k_+$ and $k_-$. In the sum of all soft-gluon attachments this piece yields
\begin{align}
\label{eq:NNLOquarkgluon}
    \!\!\xi_0 \, 6C_F \frac{1+\bar{u}_0}{\bar{u}_0^3} 
    \int_{\ell_-}^{p_{2-}} \!\frac{dk_-}{k_-}  \int^{\ell_+}_{p_{2+}} \!\frac{dk_+}{k_+}  
    \theta(k_+ k_- - m_2^2) \left(
    -2C_F\ln\frac{p_{\eta-}}{k_-} \ln\frac{\ell_+}{k_+}\right) \simeq -\frac12 \xi_0 L^4 C_F^2 \, \frac{1+\bar{u}_0}{\bar{u}_0^3} \,.
\end{align}  
Lastly, two-loop soft-quark corrections are double logarithmic if the momenta of the spectator-quark propagators are strongly ordered in their rapidity. That is, if we consider a two-loop graph of the type shown in Fig.~\ref{fig:ladderdiagrams} and label the spectator-quark momenta as $k_1$ and $k_2$ from left to right, the longitudinal momenta fulfill the conditions
\begin{align}
    \ell_- < k_{1-} < k_{2-} < p_{2-} 
     \qquad \text{and} \qquad 
     \ell_+ > k_{1+} > k_{2+} > p_{2+} \,. 
\end{align}
In that case all but the spectator-quark propagators become eikonal, the latter go on-shell and give the constraints $\theta(k_{i+} k_{i-} - m_2^2)$ for $i = 1,2$. The sum of all two-loop soft-quark contributions can then be expressed as the nested four-fold integral 
\begin{align}
\label{eq:NNLOnestedintegrals}
 \xi_0 16 C_F^2 \,\frac{1+\bar{u}_0}{\bar{u}_0^3} \, 
 \int^{p_{2-}}_{\ell_-} \! \frac{dk_{1-}}{k_{1-}} 
 \int^{p_{2-}}_{k_{1-}} \! \frac{dk_{2-}}{k_{2-}} \, \int_{m_2^2/k_{1-}}^{\ell_+} \! \frac{dk_{1+}}{k_{1+}} \int_{m_2^2/k_{2-}}^{k_{1+}} \! \frac{dk_{2+}}{k_{2+}} = \frac43 
\, \xi_0 L^4 C_F^2 \frac{1+\bar{u}_0}{\bar{u}_0^3}\,.
\end{align}
In summary, the two-loop double logarithms in the Abelian limit are given by\footnote{Note that QED corrections would lead to another source of double logarithms from the non-decoupling of soft photons from electrically charged mesons (and leptons)~\cite{Beneke:2020vnb,Beneke:2021pkl,Beneke:2022msp}. 
We do not consider those corrections here.}
\begin{align}
\label{eq:NNLODLgluequark}
F^{(2)}(\gamma) \simeq& \, \underbrace{\frac12 C_F^2 L^4 \times F^{(0)}(\gamma)}_{\text{soft-gluon corrections}} \,\, - \,\, \underbrace{C_F L^2 \times 3 \xi_0 L^2 \, C_F \frac{1+\bar{u}_0}{\bar{u}_0^3}}_{\text{soft-gluon $\times$ soft-quark}} \,\, - \,\, \underbrace{\frac12 \xi_0 L^4 \, C_F^2 \frac{1+\bar{u}_0}{\bar{u}_0^3}}_{\text{soft-quark $\otimes$ soft-gluon}} \nonumber \\[2ex]
& + \,\,  \underbrace{\frac43 \, \xi_0 L^4 C_F^2 \frac{1+\bar{u}_0}{\bar{u}_0^3}}_{\text{soft-quark corrections}} 
 \,\, = \,\,   -\xi_0 C_F^2 L^4 \, \frac{7+10\bar{u}_0}{6\bar{u}_0^3} \,,
\end{align}
where the symbol $\otimes$ implies that the soft-gluon logarithm is integrated over the longitudinal loop momenta $k_+$ and $k_-$ of the soft spectator-quark propagator. We have confirmed this result through an independent cross-check based on pole-cancellation in a bare factorization theorem derived in~\cite{Boer:2018mgl}. To this end, we have computed the leading $1/\epsilon^4$ singularities of the two-loop amplitude in the purely hard-collinear momentum region, which turns out to be the only unknown ingredient that needs to be extracted from an NNLO calculation if one focuses on the leading double logarithms. Details, together with the analysis of the full color structure, will be presented in~\cite{inprep}.

\section{Resummation of soft-quark logarithms}
\label{sec:ladders}

\begin{figure}[t]
    \centering
   \includegraphics[width=0.45\textwidth]{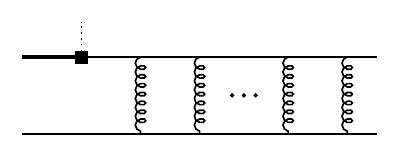}
    \caption{Diagrams that determine the Abelian soft-quark double logarithms in light-cone gauge $\bar{n} \cdot A = 0$.  
\label{fig:ladderdiagrams}}
\end{figure}

\noindent 
In this section we focus on the all-order structure of the pure soft-quark double logarithms in the Abelian limit. The diagrammatic analysis turns out to be particularly transparent in light-cone gauge $\bar{n} \cdot A = 0$, since the couplings of energetic gluons to the $b$-quark are power suppressed in this gauge. The tree-level result is therefore captured by a single diagram in this case, and the same is true for the pure (Abelian) soft-quark double logarithms at any order in perturbation theory. Specifically, these logarithms are encoded in the ladder-type diagrams  shown in Fig.~\ref{fig:ladderdiagrams}, where each loop in the ladder gives a double-logarithmic contribution if the spectator-quark propagator goes on-shell and the remaining propagators become eikonal. This is achieved by the ordering conditions
\begin{align}
    \ell_- < k_{1-} < k_{2-} < \dots < k_{n-} < p_{2-} 
    \qquad \text{and} \qquad  
    \ell_+ > k_{1+} > k_{2+} > \dots > k_{n+} > p_{2+}\,.
\end{align}
To illustrate the resummation, consider an $n$-loop generalization of the integral that appears in~\eqref{eq:NNLOnestedintegrals},
\begin{align}
\label{eq:nloopnestedintegrals}
   \int_{\ell_-}^{p_{2-}} \! \frac{dk_{1-}}{k_{1-}} \int_{k_{1-}}^{p_{2-}} \! \frac{dk_{2-}}{k_{2-}} \, \dots \int_{k_{(n-1)-}}^{p_{2-}} \! \frac{dk_{n-}}{k_{n-}}
   \, \int_{m_2^2/k_{1-}}^{\ell_+} \! \frac{dk_{1+}}{k_{1+}} \int_{m_2^2/k_{2-}}^{k_{1+}} \! \frac{dk_{2+}}{k_{2+}} \, \dots \int_{m_2^2/k_{n-}}^{k_{(n-1)+}} \! \frac{dk_{n+}}{k_{n+}} \,.
\end{align}
Multiplying this expression by a factor $\big(\frac{\alpha_s C_F}{2\pi}\big)^n$, and interpreting it as a contribution to a function $g(\ell_+,\ell_-)$ of variable light-cone momenta $\ell_+$ and $\ell_-$ gives the recursion relation\footnote{Note that the function $g(\ell_+,\ell_-)$ implicitly depends on $p_{2-}$ and $m_2$.}
\begin{align}
    g^{(n)}(\ell_+,\ell_-) = 2 C_F \int_{\ell_-}^{p_{2-}} \! \frac{dk_{-}}{k_{-}} \, \int_{m_2^2/k_{-}}^{\ell_+} \! \frac{dk_{+}}{k_{+}} \, g^{(n-1)}(k_{+},k_{-}) \,.
\end{align}
Summing over $n$ and assuming the normalization $g^{(0)}(\ell_+,\ell_-)  = 1$ leads to the integral equation
\begin{align}
\label{eq:toyintegralequation}
    \sum_{n=0}^\infty \, \Big(\frac{\alpha_s}{4\pi}\Big)^n g^{(n)}(\ell_+,\ell_-)
    = g(\ell_+,\ell_-)
    = 1 + \frac{\alpha_s C_F}{2\pi}  
    \int_{\ell_-}^{p_{2-}} \! \frac{dk_{-}}{k_{-}} \, \int_{m_2^2/k_{-}}^{\ell_+} \! \frac{dk_{+}}{k_{+}} \, g(k_{+},k_{-}) \,.
\end{align}
To resum the soft-quark logarithms to all orders, the solution to this equation needs to be evaluated for on-shell external momenta, i.e.~for $\ell_+ = \ell_- = m_2$ along with $p_{2-} = 2\gamma m_2$ in our setup. The structure is, in fact, identical to the series that determines the double-logarithmic corrections to electron-muon backward scattering at large center-of-mass energies~\cite{Gorshkov:1966qd,Bell:2022ott}. As shown in these works, the solution to the integral equation can be expressed in terms of a modified Bessel function
\begin{align} 
    g(\gamma) \equiv g(m_2,m_2) = \frac{I_1(2\sqrt{z})}{\sqrt{z}}\,, \qquad \text{with} \qquad z = \frac{\alpha_s C_F}{2\pi} L^2 \,.
\end{align}
Due to the non-trivial Dirac structure the situation is slightly more involved when considering the diagrams in Fig.~\ref{fig:ladderdiagrams}. Disentangling the numerator algebra, the soft-quark logarithms in the form factor can be described by two functions that obey a set of coupled integral equations 
\begin{align}
\label{eq:integralequations}
 f(\ell_+,\ell_-) &= 1 + \frac{\alpha_s C_F}{2\pi} \int_{\ell_-}^{p_{2-}} \!\frac{dk_-}{k_-} \int_{m_{2}^2/k_-}^{\ell_+} \!\frac{dk_+}{k_+} \, \left(f(k_+,k_-) + \frac12 f_m(k_+,k_-) \right)\,, \nonumber \\[1ex]
 f_m(\ell_+,\ell_-) &= 1 + \frac{\alpha_s C_F}{2\pi} \int_{\ell_-}^{p_{2-}} \!\frac{dk_-}{k_-} \int_{m_{2}^2/k_-}^{\ell_+} \!\frac{dk_+}{k_+} \, f_m(k_+,k_-) \,.
\end{align}
As we shall see below, the function $f_m(\ell_+,\ell_-)$ does not contribute directly to the form factor, but it mixes into $f(\ell_+,\ell_-)$ through mass insertions from the numerators of the quark propagators. Making the dependence on the mass ratio $\bar{u}_0$ explicit, we write
\begin{align}
\label{eq:FFendpointresummed}
   F(\gamma)\big\vert_\text{soft quark} &\simeq \xi_0 \left( 2\frac{1+\bar{u}_0}{\bar{u}_0^3} f(\gamma) - \frac{1}{\bar{u}_0^2} \right)\,,
   \qquad \text{with} \qquad 
   f(\gamma)\equiv f(m_2,m_2)\,.
\end{align}
The solutions to~\eqref{eq:integralequations} can again be expressed in terms of  modified Bessel functions. For the relevant function that enters~\eqref{eq:FFendpointresummed}, we find the all-order result of the pure soft-quark endpoint logarithms 
\begin{align}
   f(\gamma) = \frac12 \left( \frac{I_1(2\sqrt{z})}{\sqrt{z}} + I_0(2\sqrt{z}) \right) \,.
\end{align}
Interestingly, the Bessel functions grow exponentially in the asymptotic limit $z \to \infty$, contrary to the typical exponential Sudakov suppression. The functions $f(\ell_+,\ell_-)$ and $f_m(\ell_+,\ell_-)$ are, moreover, in direct correspondence to endpoint-divergent moments of $B_c$-meson LCDAs studied in~\cite{Boer:2018mgl}, as will be explained in more detail in~\cite{inprep}.

\section{Interplay with soft-gluon corrections}
\label{sec:sudakov}

\noindent 
To obtain the complete double-logarithmic series, soft-gluon effects need to be taken into account. As we have seen explicitly in the NNLO analysis, soft-gluon corrections modify the integrands of the nested soft-quark integrals, and therefore also the integral equations \eqref{eq:integralequations} change. Factoring out the universal Sudakov factor, we now write
\begin{align}
\label{eq:Ffinaldecimposition}
    F(\gamma) \simeq \xi_0 \,\exp\left\{ -\frac{\alpha_s C_F}{4\pi} L^2\right\} \times 
    \, \left( 2\frac{1+\bar{u}_0}{\bar{u}_0^3} f(\gamma) - \frac{1}{\bar{u}_0^2} \right) \,.
\end{align} 
The final system of integral equations, capturing all Abelian double logarithms, then takes the form
\begin{tcolorbox}[]
\vspace{-0.38cm}
\begin{align} 
\label{eq:finalintegralequations}
 f(\ell_+,\ell_-) &= 1 + \frac{\alpha_s C_F}{2\pi} \int_{\ell_-}^{p_{2-}} \!\frac{dk_-}{k_-} \int_{m_{2}^2/k_-}^{\ell_+} \!\frac{dk_+}{k_+} \, e^{-S(k_+,k_-)} \, \left(f(k_+,k_-) + \frac{1}{2} f_m(k_+,k_-) \right)\,,
 \nonumber \\[2ex]
 f_m(\ell_+,\ell_-) &= 1 + \frac{\alpha_s C_F}{2\pi} \int_{\ell_-}^{p_{2-}} \!\frac{dk_-}{k_-} \int_{m_{2}^2/k_-}^{\ell_+} \!\frac{dk_+}{k_+} \, e^{-S(k_+,k_-)} \, f_m(k_+,k_-) \,.
\end{align}
\end{tcolorbox}
\noindent
Apart from the overall factor in \eqref{eq:Ffinaldecimposition}, soft-gluon effects induce another Sudakov factor on the integrand level,
\begin{equation}
    S(k_+,k_-) = \frac{\alpha_s C_F}{2\pi} \,\ln \frac{p_{\eta-}}{k_-} \ln\frac{\ell_+}{k_+} \,,
\end{equation}
which is precisely the factor that appeared in \eqref{eq:NNLOquarkgluon}. The arguments of the respective logarithms are the ratios of the spectator-quark momentum components, $k_+$ or $k_-$, divided by the largest light-cone components that enter the loop diagram. 

Deriving a closed analytic form for the solution of the system~\eqref{eq:finalintegralequations} and the relevant function $f(\gamma)=f(m_2,m_2)$ goes beyond the scope of this article. We instead construct the first few terms in the perturbative expansion iteratively, 
\begin{align} 
    f(\gamma) = 1 + \frac32 \, \left(\frac{\alpha_s C_F}{4\pi} L^2\right) + \frac{5}{12} \, \left(\frac{\alpha_s C_F}{4\pi} L^2\right)^2 - \frac{1}{180} \, \left(\frac{\alpha_s C_F}{4\pi} L^2\right)^3 + \dots \,.
\end{align}
Inserting this expansion into~\eqref{eq:Ffinaldecimposition} reproduces the tree-level result in~\eqref{eq:xiLO}, and the NLO and NNLO results in~\eqref{eq:NLODL} (for $C_A=0$) and~\eqref{eq:NNLODLgluequark}, respectively. Interestingly, most of the features of~\eqref{eq:finalintegralequations} contribute already non-trivially at NNLO, in particular the recursive structure and the mixing pattern. To verify the exponentiation of the soft-gluon effects on the integrand level beyond the linear term, on the other hand, would require an explicit three-loop calculation. We also note that the generalization to the full non-Abelian color structure only requires minor modifications, as will be explained in~\cite{inprep}. 

Finally, we remark that this novel type of integral equations reduces to more familiar cases in certain limits. Ignoring the mixing term and dropping the Sudakov factor in the integrand leads to the double-logarithmic series of muon-electron backward-scattering, as already discussed around~\eqref{eq:toyintegralequation}. Keeping instead the Sudakov factor in the integrand, but neglecting the iterative structure of the integrals together with the tree-level contribution, gives the integral
\begin{align}
 \frac{\alpha_s C_F}{2\pi} \int_{\ell_-}^{p_{2-}} \!\frac{dk_-}{k_-} \int_{m_{2}^2/k_-}^{\ell_+} \!\frac{dk_+}{k_+} \, e^{-S(k_+,k_-)} \,.
\end{align}
For $\ell_+ = p_{2-} = m_H$ the Higgs mass, $m_{2} = m_b$ the bottom-quark mass and $\ell_-=m_b^2/m_H$,
this integral evaluates to
\begin{align}
\frac{z}{2} \, _2F_2\left(1,1;2,\frac32;-\frac{z}{4}\right)\,, \qquad \text{with} \qquad z = \frac{\alpha_s C_F}{2\pi} \ln^2 \frac{m_b^2}{m_H^2} \,,
\end{align}
and gives the double-logarithmic amplitude for the bottom-induced \mbox{$H \to \gamma \gamma$} decay~\cite{Kotsky:1997rq,Akhoury:2001mz,Liu:2018czl}. The integral equations presented in~\eqref{eq:finalintegralequations} can thus be viewed as a generalization of those cases.

\section{Conclusions}
\label{sec:conclusions}

\noindent 
We investigated the leading logarithmically-enhanced corrections to the ``soft-overlap'' contribution to $B_c \to \eta_c$ transition form factors in the non-relativistic approximation. At large hadronic recoil, double logarithms arise from a non-trivial interplay of ``endpoint'' configurations, associated with energy-ordered spectator-quark propagators, and standard exponentiated soft-gluon corrections. Their all-order structure is governed by a novel type of integral equations, and we explicitly verified that these reproduce the fixed-order double logarithms up to NNLO. Finally, we argued that the integral equations reduce to more familiar cases like muon-electron backward scattering or the bottom-induced $H \to \gamma\gamma$ decay amplitude in certain limits.

While our analysis was restricted to the Abelian limit, preliminary results suggest that the structure of the integral equations changes only slightly in full QCD~\cite{inprep}. Moreover, we stress that the iterated pattern of endpoint logarithms and their interplay with soft-gluon corrections is not a particular feature of the non-relativistic framework, but it should rather be viewed as the origin of the failure of conventional factorization techniques for power-suppressed $B$-decay amplitudes. An effective-field-theory formulation of such a nested structure of logarithms in terms of renormalization-group equations is currently unknown, but would be important to improve our understanding of soft-collinear factorization in exclusive $B$ decays. 

\subsubsection*{Acknowledgements}

\noindent 
The research of GB, TF, DH and VS was supported by the Deutsche Forschungsgemeinschaft (DFG, German Research Foundation) under grant 396021762 --- TRR 257. The work of PB was supported by the Cluster of Excellence Precision Physics, Fundamental Interactions, and Structure of Matter (PRISMA$^+$ EXC 2118/1) funded by the DFG within the German Excellence Strategy (Project ID 39083149) and by the European Research Council (ERC) under the European Union’s Horizon 2022 Research and Innovation Programme (Grant agreement No.101097780, EFT4jets). GB and PB thank the Erwin-Schrödinger International Institute for Mathematics and Physics at the University of Vienna for partial support during the Programme ``Quantum Field Theory at the Frontiers of the Strong Interactions'', July 31 - September 1, 2023.


\end{document}